\documentclass[twocolumn,preprintnumbers,amsmath,amssymb,pra]{revtex4}

\usepackage{enumerate}
\usepackage{amsmath}
\usepackage{amssymb}
\usepackage{graphicx}

\pdfoutput=1

\newcommand{\lv}{\left \vert}
\newcommand{\rv}{\right \vert}
\newcommand{\la}{\left \langle}
\newcommand{\ra}{\right \rangle}
\newcommand{\ket}[1]{\lv #1 \ra}
\newcommand{\bra}[1]{\la #1 \rv}

\begin{document}

\title{Thermal robustness of multipartite entanglement of the 1-D spin 1/2 XY model}
\author{Yoshifumi Nakata}
\affiliation{Department of Physics, Graduate School of Science, University of Tokyo, Tokyo 113-0033, Japan}

\author{Damian Markham}
\affiliation{Department of Physics, Graduate School of Science, University of Tokyo, Tokyo 113-0033, Japan\\
Universite Paris 7, 75013 Paris, France}

\author{Mio Murao}
\affiliation{Department of Physics, Graduate School of Science, University of Tokyo, Tokyo 113-0033, Japan\\
Institute for Nano Quantum Information Electronics, University of Tokyo, Tokyo 153-8505, Japan\\
PRESTO, JST, Kawaguchi, Saitama 332-0012, Japan}

 \begin{abstract}
  We study the robustness of multipartite entanglement of the ground state of the one-dimensional spin 1/2 XY model with a transverse magnetic field in the presence of thermal excitations, by investigating a threshold temperature, below which the thermal state is guaranteed to be entangled.  We obtain the threshold temperature based on the geometric measure of entanglement of the ground state.  The threshold temperature reflects three characteristic lines in the phase diagram of the correlation function. Our approach reveals a region where multipartite entanglement at zero temperature is high but is thermally fragile, and another region where multipartite entanglement at zero temperature is low but is thermally robust.
 \end{abstract}

 \date{\today}

 \maketitle
 \section{Introduction}

In quantum information, we manipulate quantum systems using unitary operations and measurements.  To achieve such manipulations with high accuracy, well-controllable quantum systems are required.  Physically, well-controllable quantum systems are likely to be described by microscopically controllable Hamiltonians.  We may describe such systems as artificial systems. On the other hand, in natural physical systems, it is rare to have a microscopically fine-controllable Hamiltonian.  Typically we cannot switch on and off individual interactions on demand and it is more likely that we only have control of a few macroscopic parameters of the Hamiltonian, such as the a magnetic field.

Entanglement is a non-local correlation that does not exist in classical mechanics, and it is considered to be a key for understanding the power of quantum computation and quantum protocols \cite{Pro}, when comparing to their classical counterparts.  Although the controllability of systems is different for artificial systems and for natural physical systems, entanglement is an indicator of the quantum signature for both systems.  Recently, investigations of quantum effects from the view point of entanglement in natural physical systems attracts increasing interest~\cite{EntPhys}.

One interesting result on the relationship between entanglement and quantum properties of natural physical systems regards entanglement of their ground state and quantum phase transitions (QPTs) \cite{QPT}.  QPTs are phase transitions at zero temperature induced by varying parameters in a Hamiltonian \cite{Sach}.  It is numerically shown in many spin models that measures of {\it bipartite} entanglement of the ground state, for example, concurrence \cite{Concurrence} of the reduced two-spin state of the ground state, or block entropy, has non-analyticity in the vicinity of the QPT points \cite{QPT}.

Since isolating the systems completely from the environment is not easy for natural physical systems where we do not have fine control of microscopic variables, it is difficult to achieve zero temperature.  Thus, it is interesting to understand the relationship between entanglement properties and quantum effects corresponding to QPT of natural physical systems at non-zero temperature. There have been several works investigating bipartite entanglement of non-zero thermal states of natural physical systems \cite{Thermal}, and for many models, it is found that concurrence and logarithmic negativity of thermal states are nonzero at finite temperature.

However, there is no guarantee that such bipartite entanglement fully describes the properties of the QPT.  In general, eigenstates are fully entangled in a {\it multipartite} manner.  From the study of multipartite entanglement in quantum information, it is known that some entanglement properties (e.g., SLOCC convertibility \cite{SLOCC}) of multipartite states cannot be characterized only by bipartite entanglement even for pure states.  For mixed states, there even exists a state where there is no bipartite entanglement between any two division, but still contains multipartite entanglement \cite{BM}. It has also been shown that resonating-valence-bond (RVB) states, which are used to describe many natural many-body systems \cite{Lee}, contain multipartite entanglement but have a negligible amount of two-site entanglement \cite{Dagomir1}.

This has led to investigations using multipartite entanglement, though this avenue is much less well developed, and there are fewer results. The main obstacle is the difficulty in calculating many of the multipartite entanglement measures for large physical systems. Due to this, one approach has been to consider multipartite entanglement witnesses, which are generally much more mathematically tractable (see e.g. \cite{Anders07,Guhne08}). This has given bounds on the minimum energy allowed for thermal states of the spin 1/2 Ising model and XX Heisenberg in a transverse magnetic field for example \cite{Brukner04,Guehne}. Also in this direction, in \cite{Dagomir2}, the authors constructed a multipartite entanglement witness for mixed systems including thermal stabilizer states related to (multiple-point) correlation functions. Another direction has been to look for entanglement measures which are easier to calculate. In \cite{Wei} the geometric measure of entanglement is studied for the ground state for the 1D spin 1/2 XY model over QPTs, and it is seen that over a phase transition, the geometric measure behaves singularly. But in general, due to the computational difficulty of analyzing mixed multipartite entanglement, multipartite entanglement in the thermal states of natural physical systems and its relationship to QPT-like behavior of the system has not been well understood.

In this paper, we study multipartite entanglement of a thermal state in the 1D spin 1/2 XY model with a transverse magnetic field to investigate the relationship between multipartite entanglement and the QPT-like behavior of natural physical systems at non-zero temperature. Instead of analyzing multipartite entanglement of an exact thermal state of the system, we investigate how thermal excitations affect the entanglement of the ground state.  We employ the threshold temperature, which is the temperature below which a thermal state is certainly entangled, defined in Ref.\cite{damian}. We derive the threshold temperature of multipartite entanglement of the ground state  measured by using a geometric measure \cite{owari} and information of population of the ground state.   This threshold temperature is an indicator to present how robust multipartite entanglement of the ground state is in the presence of thermal excitations.  In addition, we see that this approach extends the identification of borders by entanglement seen in \cite{Wei}, showing that the threshold temperature varies singularly at all of the borders of the phase diagram of the system.

This paper is composed as follows.  In Section \ref{TTSec}, we present a threshold temperature and show how an energy gap between the ground energy and the first excited energy affects to the threshold temperature.  We describe the 1D spin 1/2 XY model with a transverse magnetic field and show a phase diagram and corresponding multipartite entanglement at zero-temperature in Section \ref{XYSec}.  We present our results and analysis of the threshold temperature in the XY model in Section \ref{TTXYSec}.

\section{Threshold temperature} \label{TTSec}

\subsection{Definition}

We consider systems described by a Hamiltonian $H$ and denote the eigenenergies and eigenstates as $\{E_i\}$ and $\{ \ket{e_i^{\gamma}} \}_{\gamma=1,\cdots,d_i}$, respectively, where $i=0,1,\cdots$ and $d_i$ is the degree of degeneracy of $E_i$. By representing each eigen-subspace for $E_i$ by using a projector $P_i=\sum_{\gamma=1}^{d_i} \ket{e_i^{\gamma}} \bra{e_i^{\gamma}}$, each degenerate mixed state corresponding to $E_i$ is written by $\rho_i=P_i/d_i$.  If the system is in the thermal equilibrium at temperature $T$, the thermal state is written by $\rho_{th}(T)=\exp[-H/(k_B T)]/Z(T)$ where
$k_B$ is the Boltzmann constant and $Z(T)=\sum_i {\exp[-E_i/(k_B T)}]$ is a partition function of the system.

For multipartite systems, calculating eigenenergies and eigenstates is computationally difficult in general.  Further, even for the cases that eigenenergies and eigenstates are somehow obtained and we can explicitly write down the thermal state $\rho_{th}(T)$,
finding out whether the given thermal state $\rho_{th}(T)$ is separable or not is another computationally very difficult problem, since no simple separability criterion is known for multipartite mixed states.

Instead of using all the information of eigenenergies and eigen-subspaces of the Hamiltonian, we analyze this system by just taking into account the information of the ground state $\rho_0$ and its population $d_0 p_0$, where $p_0(T)=\exp{[-E_0/(k_B T)]/{Z(T)}}$ is the population of the ground state, to investigate multipartite entanglement of a state under thermal excitations.
Using this limited information, we will find bounds on the minimum temperature for which entanglement can be guaranteed. This is the concept of a {\it threshold temperature} introduced by \cite{damian}.

To define the threshold temperature, we first use a distance-like measure of multipartite entanglement, the robustness of entanglement \cite{robustness}.  The robustness of entanglement $R(\rho)$ for a general state $\rho$ is defined by
\begin{eqnarray} \label{robustness}
R(\rho):=\min_{\omega} \{t~|~\frac{1}{1+t}(\rho + t \omega) \in {\rm Sep} \}
\end{eqnarray}
where ${\rm Sep}$ denotes a set of all separable (mixed) states.  If we consider $(\rho + t \omega)/(1+t)$ as a mixture of a state $\rho$ with another state $\omega$ with the their populations $1/(1+t)$ and $t/(1+t)$, respectively, we can interpret $1/(1+R(\rho))$ as the population of the state $\rho$ when $\omega$ is a state which destroys entanglement of $\rho$ most effectively.

The thermal state $\rho_{th}(T)$ can be interpreted as a mixture of the ground state $\rho_{0}$ and the-rest state $\omega^\prime$ as
\begin{eqnarray}
\rho_{th} (T)=d_0 p_0 (T) \rho_0 + (1-d_0 p_0(T)) \omega^\prime
\end{eqnarray}
where $\omega^\prime=1/(1-d_0 p_0(T)) \sum_{i=1} p_i (T) \rho_i$. By definition (\ref{robustness}), the thermal state $\rho_{th} (T)$ should be entangled at the temperature $T$ if the corresponding populations satisfy
\begin{eqnarray}
p_0 (T) > \frac{1}{d_0(1+R(\rho_0))}.
\end{eqnarray}
Since the state $\omega^\prime$ is not necessarily the state $\omega$ that destroys entanglement of $\rho_0$ most effectively, this is only a sufficient, not necessary condition for entanglement.

In general, evaluation of the robustness of entanglement is still a computationally difficult problem since it requires searching the mixed state $\omega$ over all the Hilbert space.  However, it has been shown \cite{owari} that a lower bound to the robustness of entanglement is given by the geometric measure $G(\rho)$ \cite{GMEnt,owari2}
 \begin{equation}
  d[1+R(\frac{P}{d})] \geq 2^{G(\rho)} \label{GMRob},
 \end{equation}
where $P$ is the support of the state $\rho$ and $d$ is the rank of the support $P$.  The geometric measure is defined by
\begin{eqnarray}
G(\rho):=-\log_2[\max_{\left \vert \Phi \right \rangle \in {\rm Pro}} \left \langle \Phi \right \vert \rho \left \vert \Phi \right \rangle]
\end{eqnarray}
where ${\rm Pro}$ denotes a set of all product pure states.  The search of the pure product state $\ket{\Phi}$ greatly reduces the required computational efforts comparing the search of the mixed state $\omega$ for $R(\rho)$.

Although the geometric measure is a measure of entanglement for pure states, it is not for mixed states.  However, it still gives a nontrivial lower bound of the robustness of entanglement \cite{owari2}.  Therefore, we define a lower bound threshold temperature $T_{th}$ using the geometric measure which satisfies
\begin{equation}
  p_0(T_{th}) = 2^{-G(\frac{P_0}{d_0})}. \label{TT}
\end{equation}
We stress that this is a temperature below which the thermal state is {\it certainly} entangled.  It is a lower bound of a temperature for the state to be entangled and a thermal state above the threshold temperature is not necessarily separable. In this sense it fills the same role as an entanglement witness.

\subsection{Effects of the first energy gap}

Before investigating the threshold temperature $T_{th}$ of the 1-D spin 1/2 XY model, we analyze how an energy gap between the ground energy and the first excited energy affects to the threshold temperature. It is intuitively expected that a system with a smaller energy gap has a lower threshold temperature, since the population of the ground state decreases quickly by increasing temperature in such cases.

To clarify this intuition,we take into account the population of the first excited state. Our resulting relationship is that
\begin{equation}
   (1- e^{-\frac{\Delta}{k_B T}})p_0(T) > \frac{1}{d_0 [R(\rho_0)+1]}
    \Rightarrow \text{$\rho_{th}(T)$ \rm{is entangled}}, \label{oioi}
\end{equation}
where $\Delta=E_1-E_0$. The proof is given in Appendix \ref{gappedTTThm}. By using the inequality \eqref{GMRob},
we define a gapped threshold temperature $\overline{T}_{th}(\Delta)$ by $(1- e^{- \Delta/(k_B \overline{T}_{th}(\Delta))})p_0(\overline{T}_{th}(\Delta)) = 2^{-G(\frac{P_0}{d_0})}$.

This gapped threshold temperature $\overline{T}_{th}(\Delta)$ is always smaller than the corresponding threshold temperature $T_{th}$. The reason of $\overline{T}_{th}(\Delta)< T_{th}$ is that a bound of the robustness of entanglement to obtain the relationship ~\eqref{oioi} is too loose (see Appendix \ref{gappedTTThm}). We need to be aware that this does not imply that the threshold temperature decreases if we take into account the first energy gap. The gapped threshold temperature is useful to understand how the gapped threshold temperature is affected by varying the first energy gap although the gapped threshold temperature is not much use to evaluate a bound of the threshold temperature, given by Eq. ~\eqref{TT}.
In the rest of this section, we examine the gapped threshold temperature to study the effect of the first energy gap.

Since $(1- e^{-\Delta/(k_BT)}) p_0(T)>(1- e^{-\Delta'/(k_B T)})p_0(T)$ for any $T$ and any $\Delta>\Delta'$, and the geometric measure of the ground state is independent of the energy gap $\Delta$, we can conclude that $\overline{T}_{th}(\Delta)>\overline{T}_{th}(\Delta')$ for finite gaps $\Delta>\Delta'$. Thus, a smaller energy gap results in the lower gapped threshold temperature as we have expected. We present an example to display this relationship.

We consider a $m$-partite spin 1/2 system. Suppose that the ground state is a generalized W-state given by
\begin{equation}
    |W \rangle = \frac{1}{\sqrt{m}}(|\uparrow \downarrow \cdots \downarrow \rangle + |\downarrow \uparrow \cdots \downarrow \rangle+ \dots
     +|\downarrow \downarrow \cdots \uparrow \rangle)
   \end{equation}
and the population of the ground state is given by $p_0(T)=\frac{1-e^{- \delta/(k_BT)}}{1-e^{-D \delta/(k_BT)}}$ where $\delta$ is a constant and $D=2^m$ is the whole dimension of the system. We assume that the first energy gap is $\kappa \delta$. Since it has been shown that the equality in the inequality \eqref{GMRob} holds as an equality for the W-state by \cite{owari} and its geometric measure is calculated to be $(m-1) \log_2 \frac{m}{m-1}$ \cite{GMEnt} \cite{owari2}, the gapped threshold temperature $\overline{T}_{th}(\kappa \delta)$, is given by the root of the equation
\begin{equation}
    (1- e^{- \kappa \delta/(k_B\overline{T}_{th})  })(1-e^{-\delta/(k_B\overline{T}_{th})})= \frac{1}{e},
\end{equation}
in the thermodynamic limit ($m \rightarrow \infty$).

\begin{figure}[]
    \centering
    \includegraphics[width=70mm,bb=0 0 227 161]{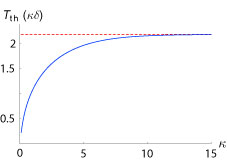}
 \caption{(Color  online) The threshold temperature $\overline{T}_{th}(\kappa \delta)$ as a function of the $\kappa$.
 We set Boltzmann constant $k_B$ as one for simplicity.
    The dashed line is the threshold temperature $T_{th}$ defined by Eq.~\eqref{TT}.
    By definition, the gapped threshold temperature $\overline{T}_{th}(\kappa \delta)$
    converges to the threshold temperature $T_{th}$
   in the limit of $\kappa \rightarrow \infty$}
    \label{TthDelta}
\end{figure}

We vary the first energy gap $\kappa \delta$ to analyze the gapped threshold temperature. Note that we fix the population of the ground state by adjusting other eigenenergies. We present the gapped threshold temperature $\overline{T}_{th}(\kappa \delta)$ as a function of the $\kappa$ in Fig. \ref{TthDelta}.  We can see that the gapped threshold temperature changes drastically with $\kappa$ in the region of small $\kappa$ although it does not change much for large $\kappa$.  This example shows that the existence of the first energy gap $\Delta$, namely whether $\Delta=0$ or $\Delta \neq 0$, is very influential on the gapped threshold temperature since the gapped threshold temperature changes drastically with $\Delta$ if the first energy gap is small.

 \section{1D spin 1/2 XY model with a transverse magnetic field} \label{XYSec}

 The dimensionless Hamiltonian of the 1D spin 1/2 XY model with a transverse magnetic field
 is given by \cite{Lieb}
 \begin{equation}
  H_{XY}=-\sum_{j=1}^{N} [\frac{1+\eta}{2}\sigma_j^X \sigma_{j+1}^X
   + \frac{1-\eta}{2}\sigma_j^Y \sigma_{j+1}^Y + h \sigma_j^Z], \label{XY1}
 \end{equation}
 where $\sigma_j^{\alpha}$ ($\alpha=X,Y,Z$) are Pauli matrices acting on a spin at $j$-th site, $\eta \in [-1,1]$ is a parameter of anisotropy and $h \geq 0$ is an external magnetic field. We consider that $\eta$ and $h$ are normalized by the coupling energy of the interaction term in the XY model Hamiltonian.  For convenience, we consider only a system with even number of particles $N$ here, but the following arguments can be similarly applied to the case of odd $N$.
 We also employ a periodic boundary condition $\sigma_{N+1}^{\alpha}:=\sigma_{1}^{\alpha}$ ($\alpha=X,Y,Z$).

 Although this XY model has been much studied in condensed matter physics, it is not so familiar in quantum information.
 We present the phase diagram of the XY model at zero temperature and the entanglement of the ground state.
 A brief explanation of the eigenstates and the eigenvalues is in Appendix \ref{XYApp}.

 \subsection{Phase diagram at zero temperature} \label{PDZ}

\begin{figure}[]
   \centering
   \includegraphics[width=70mm,bb=0 0 229 234]{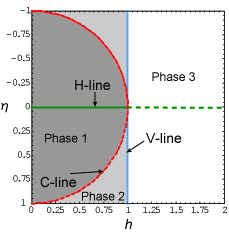}
   \caption{(Color  online) The phase diagram at zero temperature in the XY model. These phases are
   characterized by two point correlation functions. The ground states degenerate in the shaded region.
   The dashed line ($\eta=0$ and $h > 1$) in the phase diagram
   is not a phase transition line but the correlation function on this line is constant with respect to the distance $r$ similar to the C-line.}
   \label{phasefig}
  \end{figure}

 We present the phase diagram of the XY model characterized by two point correlation functions at zero temperature as in Fig.\ref{phasefig}. In different phases, two point correlation functions $C^{\alpha}(r)$ between two points separated by distance $r$, defined by $C^{\alpha}(r):=\langle \sigma_1^{\alpha} \sigma_{r+1}^{\alpha} \rangle$ for $\alpha=X,Y,Z$, have different characteristics.  Although this two point correlation functions generally decrease exponentially with the distance $r$, $C^{\alpha}(r) \sim \exp[-\frac{r}{\xi}]$ where $\xi$ is a constant called a correlation length, the rate of decrease is slower and is generally polynomial with $r$ on a border of phases.

  In the phase diagram of the XY model at zero temperature, there are three phases, phase 1, phase 2 and phase 3, which are divided by three borders referred to as a vertical line (V-line), a horizontal line (H-line) and a constant correlation line (C-line).  The phase 1 is the region of $0<h<\sqrt{1-\eta^2}$, the phase 2 is the region of $\sqrt{1-\eta^2}<h<1$ and the phase 3 is the region of $1<h$ (see Fig. \ref{phasefig}).

  The correlation functions have been calculated for large $r$ in the thermodynamic limit \cite{CF}.
  In the phase 1, the correlation functions oscillate with the distance $r$ with
  exponentially decreasing envelope. In the other regions, there is no oscillation.
  The rate of decrease in the phase 2 is different from that in the phase 3.
  Then they are considered as three different phases.
  We describe properties of the system on the V-line ($h=1$), the H-line ($\eta=0$ and $0\leq h \leq 1$)
  and on the C-line ($h^2+\eta^2=1$).

   \subsubsection{V-line and H-line}
   On the V-line and on the H-line, a quantum phase transition (QPT) occurs. QPTs are defined by the non-analyticity of the ground energy with respect to parameters in a Hamiltonian, and a line on which a $n$-th order QPT occurs is called a $n$-th order QPT line. The V-line is a second order QPT line since the second derivative of the ground energy diverges.  Due to this second order QPT, the correlation functions $C^{\alpha}(r)$ decrease only polynomially with the distance $r$ \cite{CF}.

   The H-line is a first order QPT line. The first derivative of the ground state energy diverges and there is a crossover of the eigenstates on the line for the ground state.
   This first order QPT also results in the polynomial decreasing of the correlation functions $C^{\alpha}(r)$ with the distance $r$ \cite{takahashi}.

   \subsubsection{C-line}
   The C-line is not a QPT line. However, the two point correlation functions $C^{\alpha}(r)$ are constant with respect to the distance $r$.
   In the thermodynamic limit, they are given by
 \begin{align}
  &C^X(r)=\frac{2\eta}{1+\eta},\\
  &C^Y(r)=0,\\
  &C^Z(r)=m_Z^2,
  \end{align}
  where $m_Z$ is the magnetization in the $Z$-direction defined by $m_Z=\langle \sum_{i=1}^N \sigma^Z_i/N \rangle$ \cite{CF}.
  These constant correlation functions originate from the fact that the ground state on the C-line approaches to a separable state. The explicit form of the degenerated ground states on the C-line is given by
   $|\Theta_+ \rangle =(\cos \frac{\Theta}{2} |\uparrow \rangle +\sin \frac{\Theta}{2} |\downarrow \rangle)^{\otimes N}$
   and
   $|\Theta_- \rangle =(\cos \frac{\Theta}{2} |\uparrow \rangle -\sin \frac{\Theta}{2} |\downarrow \rangle )^{\otimes N}$
   where $\cos \Theta=\sqrt{\frac{1-\eta}{1+\eta}}$ \cite{Csep}.

   Although these two ground states are both separable states, we note that the total degenerate ground state of this system not precisely separable.  Since the ground states are on the C-line degenerate, the total ground state is a mixed state given by the projector onto the subspace spanned by two ground states $| \Theta_{\pm}\rangle$,
   which is written as
   \begin{equation}
    \rho_0=\frac{1}{2}(|\Theta_+ \rangle \langle \Theta_+ |+|\Theta_+^{\perp} \rangle \langle \Theta_+^{\perp} |),
   \end{equation}
   where $|\Theta_+^{\perp} \rangle$ is a state orthogonal to $|\Theta_+ \rangle$.  The state $|\Theta_+^{\perp} \rangle$ is given by
   \begin{equation}
    |\Theta_+^{\perp}\rangle=
     \frac{1}{\sqrt{1-\cos^{2N}\Theta}}(|\Theta_-\rangle-\cos^N \Theta |\Theta_+\rangle).
   \end{equation}
   The state $|\Theta_+^{\perp} \rangle$ is easily shown to be entangled.  It is known  \cite{Ohata} that a mixture of any pure entangled state with any separable {\it pure} state is entangled. Thus, the mixed ground state $\rho_0$ is entangled. However, as we will see the next subsection, entanglement of the mixed ground state $\rho_0$ is weak and this weak entanglement results in the constant correlation functions.  In addition, we emphasize that the ground {\it energy} itself has no characteristic on the C-line.  The C-line is characterized only by properties of the ground {\it state}.

  \subsection{Geometric measure of the ground state} \label{GM}

\begin{figure}[]
   \centering
   \includegraphics[width=80mm,bb=0 0 491 371]{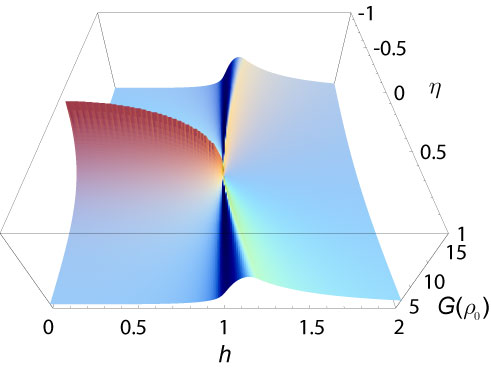}
   \caption{(Color  online) The geometric measure of the ground state of the XY model (upper) and
   its 2D plot (lower).
   The geometric measure is numerically calculated for the system of $N=80$.
   Note that the geometric measure in the region of $\eta \in [-1,0)$ is the same as that in the region of $\eta \in (0,1]$.
   The same result had been originally obtained for the region of $\eta \in [0,1]$ in Ref. \cite{Wei}.
   }
   \label{GMN80}
  \end{figure}

  We present the geometric measure of the ground state of the XY model.
  The geometric measure had been originally obtained in Ref.\cite{Wei} for the region of $\eta \in [0,1]$,
by considering the pure ground state (either $\ket{e_0^+(h,\eta)}$ or $\ket{e_0^-(h,\eta)}$ defined in Appendix \ref{XYApp})
although the ground state for $h \neq 1$ is a {\it mixed state} due to its degeneracy (see Appendix \ref{XYApp}).
In our calculation, we consider the mixed ground state $\rho_0$ for the region of $h \leq 1$. It turns out that the geometric measure of the mixed ground state coincides with that of each pure degenerated ground states in the thermodynamic limit. This is because the scaling behavior of the geometric measure of the two degenerated ground states are the same as in Ref.\cite{Wei}.  Our numerical calculation of the geometric measure of the ground state of the XY model is presented in Fig. \ref{GMN80}.

  The first derivative of the geometric measure diverges on the V-line.
  This divergence is expected to originate from the fact that the ground state depends on the variables, which
  have the non-analyticity on the V-line (see Appendix \ref{XYApp}). The non-analyticity of the variables results in the dramatic change of
  the ground state around the V-line (see Appendix \ref{tobi} for more detail).
  On the H-line, the geometric measure is non-differentiable because of the same reason.
  We note that the system in the case of $\eta=0$ has a different symmetry from other regions.

  The geometric measure on the C-line does not have distinguished characteristics to the same degree as on the V-line and the H-line.
  The geometric measure actually takes its local minimum on the C-line with respect to parameters $(h,\eta)$,
  which results in constant correlation functions (see Subsection \ref{PDZ}).
  However, since the geometric measure of the ground state is still small around the C-line and changes smoothly around the C-line.
  the C-line is difficult to identify (see the 2D plot in Fig. \ref{GM}).
  Note that the two point correlation functions are constant {\it only on the C-line}.

 \section{Threshold temperature in the XY model} \label{TTXYSec}
  \subsection{Threshold temperature and the phase diagram} \label{TTPD}
\begin{figure}[]
   \centering
   \includegraphics[width=80mm,bb=0 0 514 399]{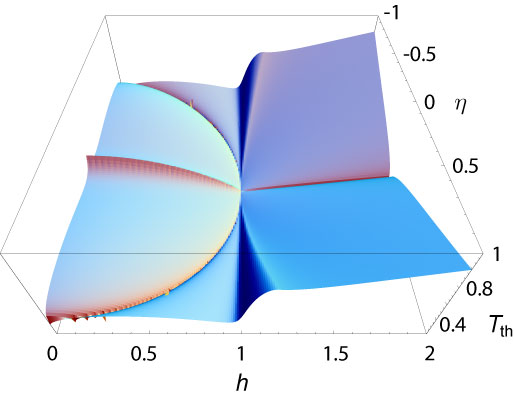}
   \caption{(Color  online) The threshold temperature in the XY model (upper) and its 2D plot (lower).
   It is calculated in the system with $N=80$. We set Boltzmann constant $k_B$ as one.
   We can clearly see the V-line, the H-line and the C-line. We note that the line of $\eta=0$, $h\geq1$ corresponds
   to the dashed line in the phase diagram (see Fig. \ref{phasefig}), which is not a QPT line.}
   \label{TTN80}
  \end{figure}

  Since the exact population of the ground state of the XY model is known \cite{takahashi}, numerical calculations of the threshold temperature is straightforward by using the values of the geometric measure of the ground state. We present the result of our numerical calculations in Fig. \ref{TTN80}.
  First, we focus on the singularity of the threshold temperature and analyze the correspondence between
  the singularity of the threshold temperature and the phase diagram at zero temperature.

  \begin{figure}[t]
  \centering
  \includegraphics[width=70mm,bb=0 0 241 169]{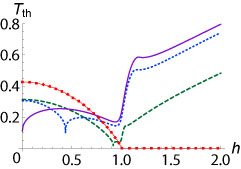}
  \caption{(Color  online) The threshold temperature shown in Fig. \ref{TTN80}. The (red) solid line with small squares, the (green) dashed line , the (blue) dotted line , and the (purple) solid line  correspond to $\eta=0, 0.4, 0.9, 1.0$, respectively.
  The threshold temperature has the singularity at $h=\sqrt{1-\eta^2}$ (on the C-line) and at $h=1$ (on the H-line).}
  \label{TT}
  \end{figure}

 On the V-line, the first derivative of the threshold temperature diverges.
 This is because the threshold temperature
 defined by $p_0(h,\eta:T_{th}) = 2^{-G(\rho_0)}$ inherits the divergence of the first derivative of the
 geometric measure on the V-line {\it and}
 the first derivative of the population of the ground state is finite.
 Similarly, on the H-line, the threshold temperature is non-differentiable.
 Its non-differentiability originates from that of the geometric measure
 of the ground state on the line.

 The threshold temperature drops dramatically on the line of $h > 1$, $\eta=0$ and on the C-line.
 These characteristics originate from
 the local minimum of the geometric measure on both lines.
 Since the ground state on the line of $h > 1$, $\eta=0$ is separable,
 which also results in constant correlation functions, the geometric measure $G(\rho_0)$
 and the threshold temperature $T_{th}$ is zero on the line.
 On the C-line, the geometric measure $G(\rho_0)$ takes local minimum with respect to the parameters $(h,\eta)$,
 which leads to a local maximum of $2^{-G(\rho_0)}$.
 Since the population of the ground state $p_0(h,\eta:T)$ decreases monotonically with temperature,
 the local maximum of $2^{-G(\rho_0)}$ yields a kink of the robustness threshold temperature
 as long as the change of the population $p_0(h,\eta:T)$ by varying $(h,\eta)$
 is less than that of the geometric measure $G(\rho_0(h,\eta:T))$.
 Thus the threshold temperature can clearly identify the
 region where entanglement of the ground state is the weakest.
 The constant correlation functions are expected to originate from weak entanglement,
 the ground state in the region where the threshold temperature dramatically drops
 will give constant correlation functions.

 We note that the shape of the singularity of the threshold temperature depends on
 the characteristics of the line. On the first order QPT line, the threshold temperature takes a kink as we can see
 on the H-line.
 On the second order QPT line, the first derivative of the threshold temperature diverges,
 which means the threshold temperature dramatically changes by varying parameters around the second order QPT line.
 We can see it on the V-line.
 Finally, if the correlation functions are approximately constant on a line, the threshold temperature
 drops dramatically on the line as shown on the C-line and on the line of $h>1$ and $\eta=0$.

 \subsection{Threshold temperature in the thermodynamic limit}

 We investigate the threshold temperature in the thermodynamic limit in this subsection. In principle, the singular behavior observed  could disappear in the thermodynamic limit, however, we find that this is not the case.

 By definition, the threshold temperature is a function of parameters $(h,\eta)$ and the system size $N$, $T_{th}(h,\eta:N)$.
 In this subsection, we show that the threshold temperature for large $N$ is independent of the system size $N$, namely,
 $T_{th}(h,\eta:N)=T_{th}(h,\eta)$ for large $N$.
 It is numerically confirmed that $N=80$ is large enough for achieving the thermodynamic limit for this system.

 The proof of $T_{th}(h,\eta:N)=T_{th}(h,\eta)$ is based on the following two facts.
 First, the geometric measure of the ground state of the XY model
 scales polynomially with $N$ for large $N$, $G(\rho_0) \propto \zeta(h,\eta) N$
 where a function $\zeta(h, \eta)$ is determined independently of $N$.
 Second, in large systems, for any $\mu \in [0,1]$ independent of $N$,
 there exists an unique temperature $\tilde{T}(\mu)$ such that
 $p_0(h,\eta:\tilde{T}(\mu):N) = 2^{-\mu N}$.
 To show that, we prove that the scaling of the population of the ground state
 $p_0(h,\eta:T:N)$ is exponential with the system size $N$, $p_0(h,\eta:T:N)\propto 2^{-\mu(h,\eta:T)N}$
 where $\log_2(1+e^{-(h+1)/(k_BT)}) \leq \mu(h,\eta:T) \leq \log_2(1+e^{-\delta(h,\eta)/(k_BT)})$
 where $d(h,\eta)=|h-1|$ for $h \geq 1-\eta^2$ and
 $d(h,\eta)=\eta \sqrt{1-\eta^2-h^2}/\sqrt{1-\eta^2}$ for $h < 1-\eta^2$.
 Since the exponent $\mu(h,\eta:T)$ is monotonically increasing with temperature,
 the statement holds.

 From the first fact, the definition of the threshold temperature, $p_0(h,\eta:T_{th}:N) = 2^{-G(\rho_0)}$,
 is rewritten as
 \begin{equation}
  p_0(h,\eta:T_{th}(h,\eta:N):N)=2^{-\zeta(h,\eta)N}
 \end{equation}
 for large $N$.
 On the other hand, due to the second fact,
 there exists an unique temperature $\tilde{T}(\zeta(h,\eta))$ such that
 \begin{equation}
  p_0(h,\eta:\tilde{T}(\zeta(h,\eta)):N)=2^{-\zeta(h,\eta) N}
 \end{equation}
 if $N$ is large enough.
 Due to these two equations and the fact that the population of the ground state
 is a single-valued function with respect to temperature,
 we can conclude that $T_{th}(h,\eta:N)= \tilde{T}(\zeta(h,\eta))$ in large systems, namely the threshold temperature is independent of $N$ for large $N$.

\section{Thermal robustness of entanglement of the ground state}

We now look in more detail at the value of the threshold temperature, beyond the singular behavior of the threshold temperature.
In the study of the threshold temperature related to the phase diagram at zero temperature, it is enough to analyze its singularity
since the phase diagram is also determined by the singular behavior of the correlation functions,
in which quantitative differences are negligible.

On the other hand, we can study the threshold temperature as an indicator of the thermal robustness of entanglement of the ground state
by investigating its value.
It may be natural to think that the stronger (weaker) the entanglement of the ground state, the higher (lower) the temperature at which the entanglement can survive in a thermal state.
However, this is not always correct.
In this section, we describe the characteristic behaviors of multipartite entanglement caused by the existence of thermal excitations in two regions, one is on the H-line, and the other is for $h \gg 1$. In those regions, we have find this natural intuition is broken.

 \subsection{Threshold temperature on the H-line}

 In Fig. \ref{GMN80}, the geometric measure of the ground state takes the maximum on the H-line.
 This fact results in the local maximum of the threshold temperature on the line, as we explained in Subsection \ref{TTPD}.
 But the peak of the threshold temperature on the line is less emphasized compared to that of the
 geometric measure.
 As long as the population of the ground state does not change so much with parameters,
 the maximum of the geometric measure is inherited by the threshold temperature, and the peak of the maximum of the threshold temperature should be more pronounced than that of the geometric measure.
 However, on the H-line, the population of the ground state takes local minimum since the first excited energy gap, between the ground energy and the first excited energy, approaches zero and the system reduces to a gapless system in the thermodynamic limit.
 This local minimum of the population of the ground state leads a smoother peak of the threshold temperature on the line.

 Therefore, there is a balance between the enhancement of the peak due to the local maximum of entanglement of the ground state and the smoothing of the peak due to the local minimum of the population of the ground state.
 As a net result, the peak of the threshold temperature is not enhanced much. Thus, multipartite entanglement at zero temperature is high on the H-line, but is fragile against thermal excitations.

 \subsection{Threshold temperature in the region of $h \gg 1$}

 \begin{figure}[h]
  \centering
  \includegraphics[width=70mm,bb=0 0 241 176]{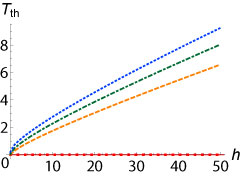}
  \caption{(Color  online) The threshold temperature in the XY model calculated in the system with $N=80$.
  We set Boltzmann constant $k_B$ as one.
  The (red) solid line with small squares, the (orange) dashed line, the (green) dotted dashed line and the (blue) dotted line show the threshold temperature
  for $\eta=0, 0.1, 0.4, 0.9$, respectively.
   The parameter $h$ is between $0$ and $50$. We can see that the threshold temperature is monotonically
   increasing with $h$ in the region of $h \gg 1$.
  We note that the threshold temperature for the case of $\eta=0$ and $h \geq 1$ is zero since the ground state is
  separable in that case.
  We can see that the effect of nonzero  $\eta$ to the threshold temperature is high although
  entanglement itself is small (see Fig. \ref{GMN80}).}
  \label{hTT}
 \end{figure}

 We now look at the threshold temperature in the XY model for $h \gg 1$.
 On the line of $h \geq 1$ and $\eta=0$ the ground state is separable and,
 in the region of $h\geq1$ and $\eta \neq 0$, the geometric measure of the entanglement of the ground state
 has small but nonzero value due to the effect of nonzero $\eta$.
 The threshold temperature is given as in Fig. \ref{hTT}.
 We present only for the case of $\eta \in [0,1]$.

 We can see that the threshold temperature increases monotonically with the external magnetic field $h$
 in the region of $h \gg 1$ for $\eta \neq 0$.
 Since a thermal state below the threshold temperature is guaranteed to be entangled,
 a thermal state of the XY model for $h \gg 1$ is entangled even at high temperature,
 although the geometric measure of the ground state is monotonically decreasing in that region.
 It may be counterintuitive that such weak entanglement of the ground state can survive
 in a thermal state at high temperature.
 The high threshold temperature originates from the fact that
 the first energy gap between the ground energy and the first excited
 energy increases more rapidly than the rate of decrease of the geometric measure.
 Thus weak entanglement of the ground state does not necessarily imply that it
 is easily destroyed by thermal excitations.

 We note that the high threshold temperature does not mean thermal effects can boost the amount of the entanglement. The ground state in the region of $h \gg 1$ for $\eta \neq 0$ is very close to a product state and the amount of entanglement is infinitesimally small but entanglement is thermally robust in a sense that it can survive until high temperature.
 In relatively artificial systems considered in quantum information, it has been found \cite{Eisert} that a multipartite entangled state which is close to a product state can be useful as a universal resource for the measurement based quantum computation \cite{Raussendorf}.
In this case, the existence of infinitesimal entanglement provides a significant difference from zero-entanglement cases.
Also, a related approach on entanglement to that presented here provides an entanglement witness very closely linked to the threshold
temperature which manifest themselves as physical observable in some spin systems \cite{Brukner06}.
However, so far, we do not know whether such an infinitesimal amount but thermally robust entanglement can contribute to present any quantum property of natural physical systems at finite temperature.  It is desirable to investigate physical properties at finite temperature related to entanglement of the ground state in the region of $h \gg 1$.  On the other hand, even if it would turn out that the infinitesimal amount of entanglement does not contribute to quantum properties at the limit of $h \gg 1$, it is interesting to investigate how multipartite entanglement at finite temperature contributes to quantum properties of the system in the trade-off of thermal robustness of entanglement and the amount of entanglement of the ground state in the intermediate region $h>1$.

 \section{Conclusion}

 In this paper, we have studied thermal robustness of multipartite entanglement of the ground state of the one-dimensional spin 1/2 XY model with a transverse magnetic field by using the threshold temperature introduced by \cite{damian}.  The threshold temperature is a temperature where the thermal state of a multipartite system is certainly entangled and its lower bound of the temperature can be obtained in terms of the geometric measure.  We have calculated the geometric measure and threshold temperature in the spin 1/2 one dimensional XY model. We have analyzed
the correspondence between the threshold temperature and the phase diagram at zero temperature by looking at the singularity of the threshold temperature. We have also analyzed thermal robustness of multipartite entanglement of the ground state.

 It has been shown that the C-line, on which two point correlation functions are constant and entanglement of the ground state takes local minimum with respect to parameters in the Hamiltonian, is clearly reflected by the threshold temperature as well as the V-line (the second order QPT line) and the H-line (the first order QPT line).  Although the geometric measure of the ground state itself does not reflect the C-line clearly, the C-line is enhanced by taking into account thermal excitations.  We also present a region in the phase diagram where multipartite entanglement at zero temperature is high but is thermally fragile, and another region where multipartite entanglement at zero temperature is less but is thermally robust.  These entanglement properties cannot be deduced only from analyzing the ground state. They provide novel information of the quantum effects corresponding to QPT for natural physical systems at the non-zero temperature.

 \section*{Acknowledgments}
 We acknowledge financial support by Special Coordination Funds for Promoting Science
 and Technology, by JST Strategic Japanese-French Cooperative Program and by QICS.

 \appendix
 \section{Gapped threshold temperature} \label{gappedTTThm}
 We show that
 \begin{eqnarray}
  (1- e^{-\Delta/(k_BT)})p_0(T) > \frac{1}{d_0 [R(\rho_0)+1]} \nonumber \\
   \Rightarrow \text{$\rho_{th}(T)$ \rm{is entangled}}.
 \end{eqnarray}
 Let $\rho_i:=P_i/d_i$ be a mixed state for the $i$-th excitation.
 A thermal state $\rho_{th}(T)$ can be rewritten as
   \begin{align}
    \rho_{th}(T)=(&p_0(T)d_0 + p_1(T)d_1) \notag \\
    &\times [\rho_{01}+(\frac{1}{p_0(T)d_0 + p_1(T)d_1}-1)\rho_{rest}],
   \end{align}
   where $p_i(T)=\exp[-E_i/(k_BT)]/Z(T)$ and
   \begin{align}
    &\rho_{01}=\frac{p_0(T)d_0 \rho_0 + p_1(T) d_1 \rho_1}{p_0(T)d_0 + p_1(T)d_1}\\
    &\rho_{rest}=\frac{1}{1-(p_0(T)d_0 + p_1(T)d_1)}\sum_{i\neq 0,1}p_i d_i \rho_i.
   \end{align}
   By definition of the robustness of entanglement,
   \begin{equation}
   \frac{1}{p_0(T)d_0 + p_1(T)d_1}-1< R(\rho_{01}) \Rightarrow \text{$\rho_{th}(T)$ \rm{is entangled}}. \label{hoi}
   \end{equation}
   We relate the robustness of entanglement of the state $\rho_{01}$, $R(\rho_{01})$,
   to that of the ground state, $R(\rho_0)$.
   The state $\rho_{01}$ can be divided into the ground state $\rho_0$ and the another state as
   \begin{eqnarray}
    \rho_{01}&=&\frac{(1-e^{-\Delta/(k_BT)})d_0}{d_0+e^{-\Delta/(k_BT)}d_1}
    \rho_0 \nonumber \\
   &+& \frac{(d_0+d_1)e^{-\Delta/(k_BT)}}{d_0+e^{-\Delta/(k_BT)}d_1}\frac{d_0 \rho_0+d_1 \rho_1}{d_0+d_1}.
   \end{eqnarray}
   Since the robustness of entanglement satisfies
   \begin{equation}
    R(p \sigma+(1-p) \sigma')+1 \geq p[R(\sigma)+1] \label{tda}
   \end{equation}
   for $p \geq 0$, $p \in[0,1]$ and any states $\sigma, \sigma'$.
   the robustness $R(\rho_{01})$ and $R(\rho_0)$ can be related as
   \begin{equation}
    R(\rho_{01})+1 \geq \frac{(1-e^{-\Delta/(k_BT)})d_0}{d_0+e^{-\Delta/(k_BT)}d_1} [R(\rho_0)+1].
   \end{equation}
   By replacing the right-hand side of the inequality in Eq.~\eqref{hoi},
   we can obtain the inequality that
   \begin{equation}
    (1- e^{-\Delta/(k_BT)})p_0(T)  > \frac{1}{d_0 [R(\rho_0)+1]}.
   \end{equation}

   The reason why the gapped threshold temperature is always smaller than the threshold temperature
   is that the bound of the robustness of entanglement ~\eqref{tda} is not tight.
   In general, the threshold temperature is expected to be improved by taking into account of the first energy gap as we can see from the proof in Ref. \cite{damian}.

We also note that it is possible to bound $p_0$ simply in terms of the gap $\Delta$, as was done in \cite{damian}, however, in general this is not useful since it gives very low threshold temperatures, going to zero in the thermodynamic limit.

\section{Eigenstates and eigenenergy of the XY model} \label{XYApp}

We present a brief explanation of the XY model. The Hamiltonian is given by
 \begin{equation}
  H_{XY}=-\sum_{j=1}^{N} [\frac{1+\eta}{2}\sigma_j^X \sigma_{j+1}^X
   + \frac{1-\eta}{2}\sigma_j^Y \sigma_{j+1}^Y + h \sigma_j^Z].
 \end{equation}
 Although we consider $\eta \in [-1,1]$, it is enough to study the case of $\eta \in [0,1]$. This is because the Hamiltonian in the region of $\eta \in [-1,0)$, say $H_{XY}^{[-1,0)}$, can be obtained by multiplying
 a local unitary by the Hamiltonian in the region of $\eta \in (0,1]$, say $H_{XY}^{(0,1]}$, namely
 \begin{equation}
  H_{XY}^{[-1,0)}=(\otimes_{i=1}^N U_i)  H_{XY}^{(0,1]} (\otimes_{i=1}^N U_i^{\dagger}).
 \end{equation}
 where $U_i=(I+i \sigma^Z_i)/\sqrt{2}$.
 Hence we can easily construct the eigenstates and the eigenenergy of the Hamiltonian $H_{XY}^{[-1,0)}$ from those
 of the Hamiltonian $H_{XY}^{(0,1]}$.
 Similarly, we can extend the region of parameters to arbitrary $h$ by multiplying by another
 local unitary.  In the following explanation of the XY model, we consider only the case of $\eta \in [0,1]$.

 This Hamiltonian is exactly diagonalized by
 the Jordan-Wigner transformation $c_k:=\otimes_{j=1}^{k-1} (-\sigma_j^Z) \sigma_k^-$
 where $\sigma^{\pm}_k:=(\sigma_k^X \pm i\sigma_k^Y)/\sqrt{2}$,
 the Fourier transformation $f_k:=\frac{1}{\sqrt{N}} \sum_{j=1}^{N}\exp[-(\theta_k^{\pm} j+\frac{\pi}{4})i]c_j$
 and the Bogoliubov transformation $\Gamma_k^{\pm}(h,\eta):=\cos \phi_k^{\pm}(h,\eta) f_k -\sin \phi_k^{\pm}(h,\eta) f_{-k}^{\pm \dagger}$,
 where $\theta^+_k=\frac{(2k-1) \pi }{N}$, $\theta^-_k=\frac{2k \pi }{N}$ and $\{\phi_k^{\pm}(h,\eta)\}$ are
 given by
 \begin{eqnarray}
  &\eta \sin \theta^{\pm}_k  \cos 2 \phi^{\pm}_k + (h+\cos \theta^{\pm}_k) \sin2\phi^{\pm}_k=0,& \label{phidef1}\\
  &\eta \sin \theta^{\pm}_k  \sin 2 \phi^{\pm}_k - (h+\cos \theta^{\pm}_k) \cos2\phi^{\pm}_k \geq 0.\label{phidef2}&
 \end{eqnarray}

 Since the Hamiltonian $H_{XY}$ commutes with parity operator $\otimes_j(1-2\sigma_j^+ \sigma_j^-)$,
 we can divide the whole Hilbert space $\mathcal{H}_{XY}$ into a parity plus (minus) subspace, $\mathcal{H}_{XY}^{\pm}$, as
 \begin{equation}
  \mathcal{H}_{XY}=\mathcal{H}_{XY}^+ \oplus \mathcal{H}_{XY}^-.
 \end{equation}
 The states in a parity plus (minus) subspace have even (odd) number of up spins.
 The Hamiltonians $H_{XY}^{\pm}$ projected onto each subspace $\mathcal{H}_{XY}^{\pm}$ are given by
 \begin{equation}
  H_{XY}^{\pm}= \sum_{k=1}^N \varepsilon^{\pm}_k(h,\eta)(\Gamma^{\pm \dagger}_k(h,\eta)
   \Gamma^{\pm}_k(h,\eta)-\frac{1}{2}),
 \end{equation}
 where $\varepsilon^{\pm}(k):=2\sqrt{(h+\cos \theta^{\pm}_k)^2+ \eta^2  \sin^2 \theta^{\pm}_k}( \geq 0)$.
 Since the operators $\Gamma^{\pm \dagger}_k$ ($k=1,\cdots,N$) satisfy the anti-commutation relation,
 they can be regarded as generation operators of pseudo fermions with an energy $\varepsilon^{\pm}(k)$.
 The lowest energy for each Hamiltonian, $H_{XY}^{+}$ or $H_{XY}^{-}$, is given by
 \begin{align}
  E_0^+(h,\eta)&=-\sum_{k=1}^N \sqrt{(h+\cos \theta^+_k)^2+ \eta^2  \sin^2 \theta^+_k}, \\
  E_0^-(h,\eta)&=-\Xi(h)-\sum_{k=1}^N \sqrt{(h+\cos \theta^-_k)^2+ \eta^2  \sin^2 \theta^-_k},
 \end{align}
 respectively, where $\Xi(h)=0$ for $h\leq 1$ and $\Xi(h)=2(h-1)$ for $h> 1$.
 The eigenstates corresponding to the lowest energy in each subspaces are given by
 \begin{align}
  &|e_0^+\rangle=\prod_{k=1}^{\frac{N}{2}}(\cos \phi^+_k + \sin \phi^+_k
  f^{+ \dagger}_k f^{+ \dagger}_{-k}) |\downarrow \cdots \downarrow \rangle, \label{G1} \\
  &|e_0^-\rangle=f^{- \dagger}_N \prod_{k=1}^{\frac{N}{2}-1}(\cos \phi^-_k + \sin \phi^-_k
  f^{- \dagger}_k f^{- \dagger}_{-k}) |\downarrow \cdots \downarrow \rangle. \label{G2}
 \end{align}

 In the region of $h >1$, the ground energy $E_0(h,\eta)$ is equal to $E_0^+(h,\eta)$ and the ground state is $|e_0^+\rangle$.
 In the region of $h \leq 1$, two energies $E_0^{\pm}(h,\eta)$ coincide in the thermodynamic limit ($N\rightarrow \infty$)
 and the ground states degenerate.
 Since a thermal state at zero temperature reduces to
 a projector onto the subspace spanned by the degenerated ground state,
 the ground state in the thermodynamic limit in the region of $h \leq 1$ is a mixed state.

 The eigenstates $|e_0^{\pm}\rangle$ in Eq.~\eqref{G1} and Eq.~\eqref{G2}
 are represented by global operators $\{f_k^{\pm} \}_k$.
 To investigate entanglement of the states, the states should be represented in a local basis. Thus we have to transform global operators back to local operators defined at each site.

 \section{Investigation of $\phi^{\pm}(h,\eta)$ on the V-line and on the H-line} \label{tobi}
  \begin{figure}[]
  \centering
   \includegraphics[width=70mm, bb=0 0 294 192]{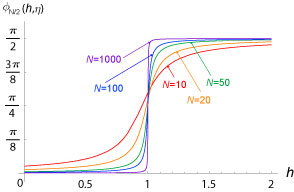}
  \includegraphics[width=70mm, bb=0 0 1500 928]{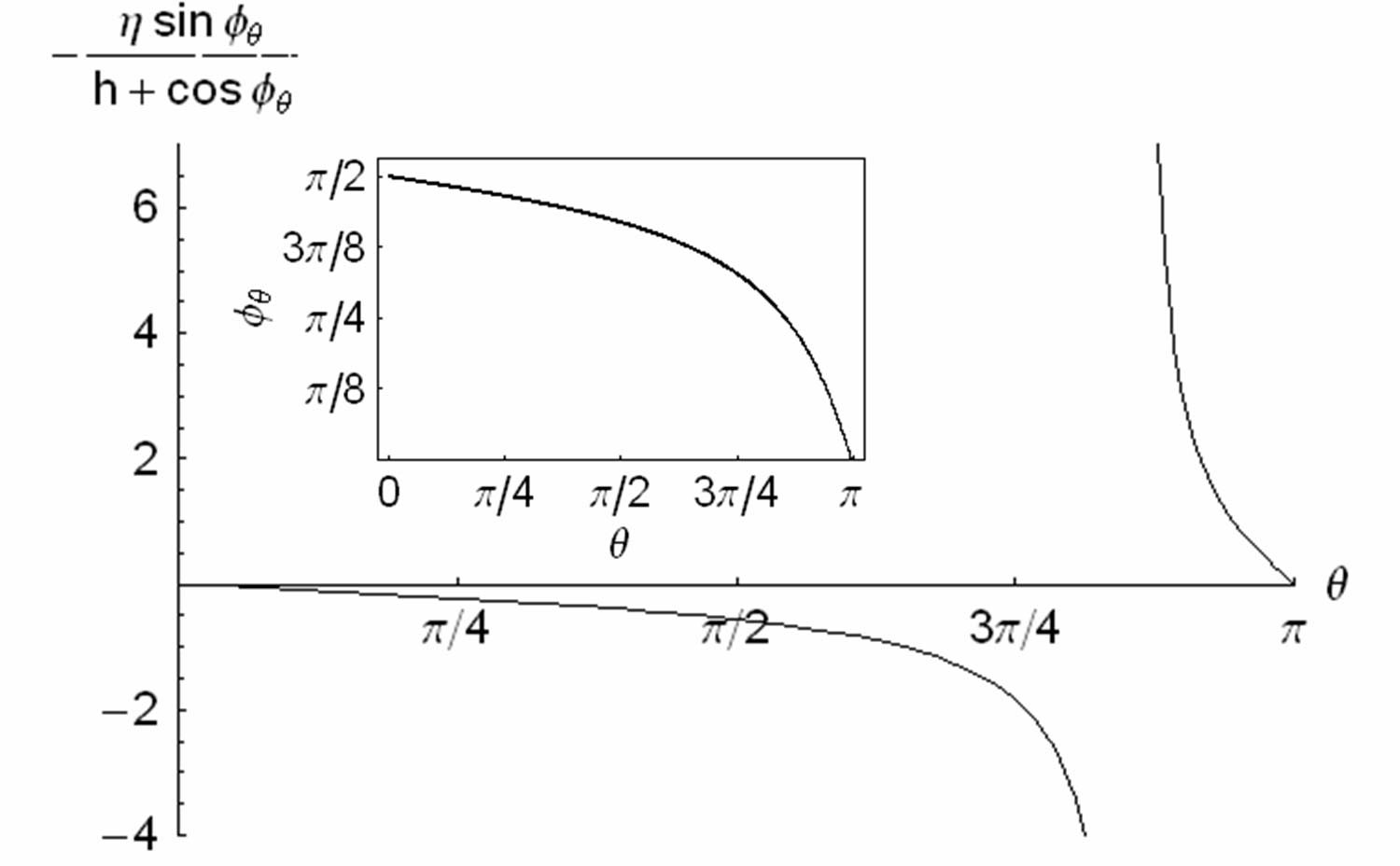}
  \includegraphics[width=70mm, bb=0 0 1500 927]{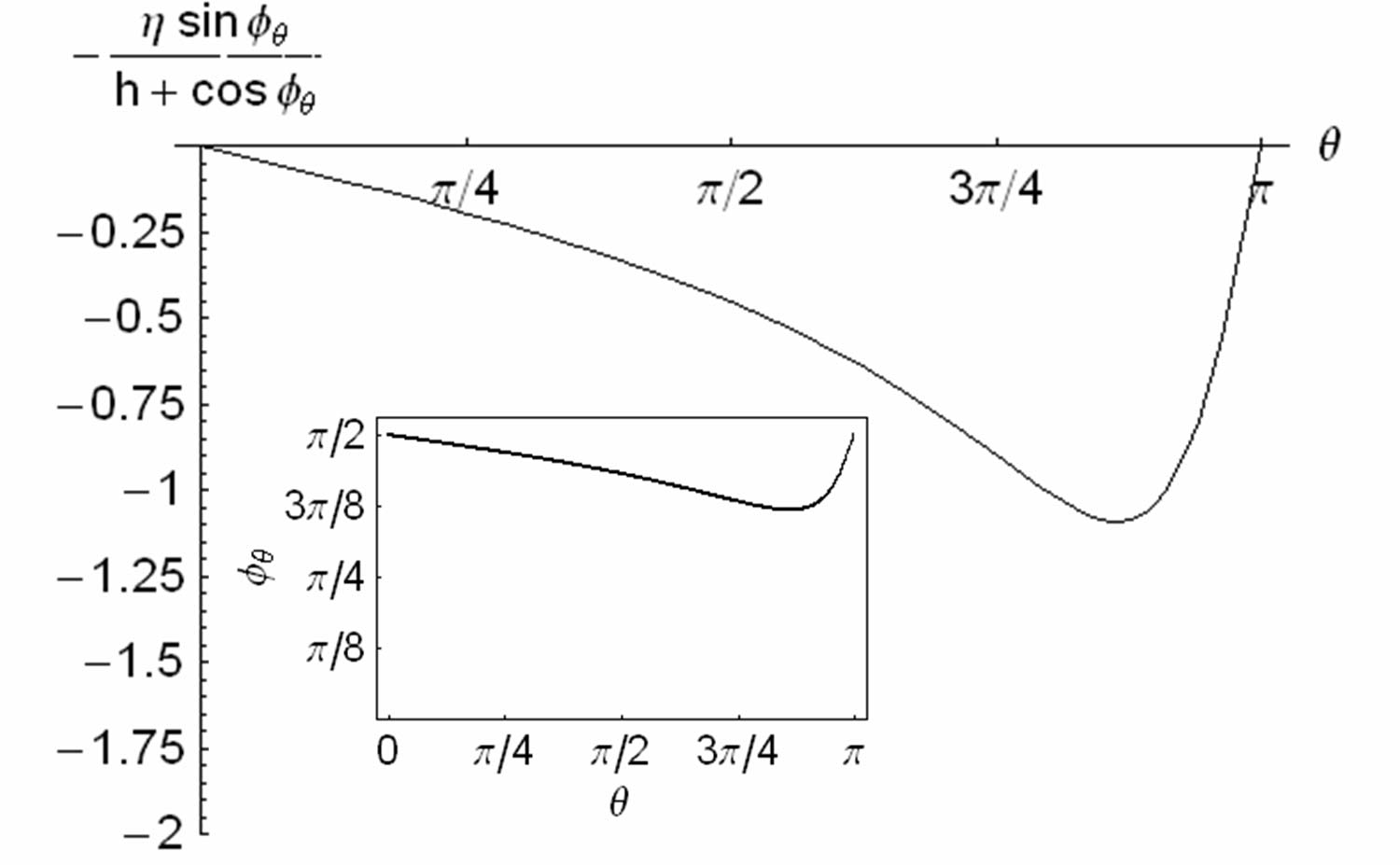}
  \caption{
   (Color  online) The upper figure is the numerical calculations of $\phi_{N/2}^{+}(h,0.5)$ for $N=10,20,50,100,1000$. We can see the non-analyticity at $h=1$.
   The middle figure and the lower figure are
   the right-hand side of Eq.~\eqref{defth} in the case of $h=0.9$ (middle) and $h=1.1$ (lower)
   with respect to $\theta$. We fixed $\eta=0.5$.
   The inset is $\phi_{\theta}(h,\eta)$ with respect to $\theta$ respectively.}
  \label{tantan}
 \end{figure}
 We show that
 $\langle e_0^{\pm}(1-\delta h,\eta)| e_0^{\pm}(1+\delta h , \eta) \rangle=0$ for any small $\delta h>0$ in the thermodynamic limit, which had been numerically obtained in Ref. \cite{jump}
 For that, we first investigate the variables $\{\phi_k^{\pm}(h,\eta)\}_k$ defined by Eq.~\eqref{phidef1}
 and the inequality ~\eqref{phidef2}, which are rewritten as
 \begin{align}
  \tan 2 \phi_{k}^{\pm}(h,\eta)&=-\frac{\eta \sin \theta_k^{\pm}}{h+\cos \theta_k^{\pm}}, \\
  \sin 2 \phi_{k}^{\pm}(h,\eta) &\geq 0.
 \end{align}
 Since $\theta^+_k=\frac{(2k-1) \pi }{N}$, $\theta^-_k=\frac{2k \pi }{N}$, the variables $\{\phi_k^{\pm}(h,\eta)\}_k$ reduces to $\{\phi_{\theta}(h,\eta)\}_{\theta \in (0, \pi)}$, in the thermodynamic limit, defined by
 \begin{align}
  \tan 2 \phi_{\theta}(h,\eta)&=-\frac{\eta \sin \theta}{h+\cos \theta}, \label{defth}\\
  \sin 2 \phi_{\theta}(h,\eta) &\geq 0.
 \end{align}

 We show that, for any small $\delta h>0$,
 \begin{align}
  \lim_{\theta \rightarrow \pi} \phi_{\theta}(1-\delta h,\eta)&=0, \label{hsmaller}\\
  \lim_{\theta \rightarrow \pi} \phi_{\theta}(1+\delta h,\eta)&=\frac{\pi}{2}. \label{hlarger}
 \end{align}
 In the case of $h=1-\delta h$, there exists $\theta_0$ such that
 the right-hand side of Eq.~\eqref{defth} diverges.
 Then $\tan 2\phi_{\theta}(1-\delta h,\eta) > 0$ for $\arccos (-h) < \theta < \pi$
 (see Fig. \ref{tantan}).
 Therefore $\lim_{\theta \rightarrow \pi} \phi_{\theta}(1-\delta h,\eta)=0$.
 On the other hand, in the case of $h=1+\delta h$, the right-hand side of Eq.~\eqref{defth} is always
 finite, which implies $\tan 2\phi_{\theta}(1-\delta h,\eta) < 0$ for any $\theta \in (0,\pi)$
 (see Fig. \ref{tantan}).
 Then $\lim_{\theta \rightarrow \pi}  \phi_{\theta}(1+\delta h,\eta)=\frac{\pi}{2}$.

 Due to this fact, we can easily show that $\langle e_0^{\pm}(1-\delta h,\eta)| e_0^{\pm}(1+\delta h,\eta) \rangle=0$
 for any small $\delta h>0$ in the thermodynamic limit.
 The inner product can be rewritten as
 \begin{eqnarray}
  \langle e_0^{\pm}(1-\delta h,\eta)| e_0^{\pm}(1+\delta h,\eta) \rangle \nonumber \\
   = \prod_{k=1}^{\mathcal{N}}\cos[\phi^{\pm}_k(1+\delta h) -\phi^{\pm}_k(1-\delta h)],
 \end{eqnarray}
 where $\mathcal{N}$ is $N/2$ for the plus case and $N/2-1$ for the minus case.
 Since Eq.~\eqref{hsmaller} and Eq.~\eqref{hlarger} imply
 $\phi_{\mathcal{N}}^{\pm}(1+\delta h,\eta) - \phi_{\mathcal{N}}^{\pm}(1-\delta h,\eta) = \frac{\pi}{2}$ in
 the thermodynamic limit, the inner product is zero.
 As long as no accidental cancelation of terms occurs in the calculation of the geometric measure,
 the geometric measure inherits this abrupt change of the ground state and has non-analyticity on the V-line as explained in Subsection \ref{GM}.

 In a similar way, it is proven that
 $\langle e_0^{\pm}(h,\delta \eta)| e_0^{\pm}(h,-\delta \eta) \rangle=0$ for any small $\delta \eta>0$
 in the thermodynamic limit. Then the ground state also has an dramatic change around the H-line.

 We note that, since the variables $\{\phi_{\theta}(h,\eta)\}$ completely determine
 how the ground state depends on the parameters $(h,\eta)$,
 the non-analyticity of these variables will inherit to the ground state.

\end{document}